\documentstyle[eqsecnum,preprint,prd,aps]{revtex}
\begin{document}
 \thispagestyle{empty}
\draft
\title{W, Z and Higgs Scattering at SSC Energies}
\author{Suraj N. Gupta and James M. Johnson}
\address{ Department of Physics, Wayne State University, Detroit,
	Michigan 48202}
\author{Wayne W. Repko}
\address{Department of Physics and Astronomy, Michigan State University,
	East Lansing, Michigan 48824}
\date{March 3, 1993}
\maketitle
\begin{abstract}
	The scattering of $W$, $Z$ and Higgs bosons in the Standard
Model is investigated in the region $s,m_H^2\gg m_W^2$ with no
restrictions on relative sizes of $s$ and $m_H^2$, so that our results
are applicable at energies below as well as above $2m_H$.  We have
calculated, with the inclusion of the full one-loop corrections, the
scattering matrix between the states $W_L^+W_L^-$, $Z_LZ_L$ and $HH$,
and computed the S-wave amplitudes as functions of the center-of-mass
energy $\sqrt{s}$ for $m_H=$500~GeV and 1000~GeV.  The apparent
violation of unitarity is avoided by unitarizing the amplitudes by the
K-matrix and the Pad\'e methods.  For the detection of the Higgs boson
through gauge boson scattering in $pp$ collisions, we have used the
unitarized amplitudes to obtain the invariant-mass distributions for
the final $W_L^+W_L^-$ and $Z_LZ_L$ pairs at the SSC energy of
$\sqrt{s}=$40~TeV by means of the effective-W approximation.
\end{abstract}

\narrowtext
\raggedright
\setlength{\parindent}{24pt}
\section{INTRODUCTION}

	Since the Higgs boson, which plays an essential role in the
Standard Model of electroweak interactions, has proved elusive
so far, it is of utmost interest to explore phenomena such as the $W$,
$Z$ and Higgs scattering at SSC energies.  It is particularly
interesting to consider the region $s,m_H^2\gg m_W^2$, which allows the
use of the Goldstone-boson equivalence theorem\cite{LQT,equiv}. More
importantly, in this region the weak interactions no longer remain
weak, and we may expect to observe striking effects in scattering
processes.

	At high energies the tree amplitudes are theoretically
inadequate to describe the scattering of gauge bosons, and it
is necessary to extend the earlier tree level results\cite{LQT} at
least to one loop.  There has been a flurry of activity in recent years
to deal with this problem, and several authors
\cite{DW,dicus,veltman,passarino,DJM}
have covered the complete treatment of the $W$
and $Z$ scattering to one loop at energies below $2m_H$. We shall now
provide the treatment at energies above $2m_H$, which requires the
inclusion of the $HH$ channel\cite{DJLnote}. It should be noted that while the
one-loop calculations of Refs.~\cite{DW} and~\cite{dicus} remain valid at
all energies, they do not include the effect of the $HH$ channel on
$W$ and $Z$ scattering at energies above $2m_H$.

	The investigation of the gauge-boson scattering at energies
above $2m_H$ represents a major theoretical undertaking because
many additional one-loop diagrams have to be evaluated due to the
opening of the $HH$~channel. Moreover, an SO(3) symmetry, which was
used to simplify the calculations in the earlier papers, cannot be
extended to the $HH$ channel. The analytic evaluation of triangle and
box diagrams for the $HH$ channel is also more challenging than the
evaluation of similar diagrams in the earlier investigations.  We have
overcome these complexities and obtained the numerous real and
imaginary terms arising from the new diagrams.

	During the early stages of the development of the covariant
perturbation theory, it was pointed out\cite{gupta} that
although Feynman's expansion of the scattering operator is unitary as a
whole, it violates unitarity up to any finite order, and it was argued
that unitarization of the scattering operator should be achieved by
means of the K-matrix expansion.  Subsequently, some other authors
advocated the use of a different unitarization procedure known as the
Pad\'e expansion\cite{pade}. Both the K-matrix and the Pad\'e
unitarizations have been applied to the $W$ and $Z$ scattering at energies
below $2m_H$\cite{cj,dicus,dobado}, and we shall explore the energy
region above $2m_H$ with the inclusion of the $HH$ channel.
As shown in Ref.~\cite{DW}, it is also possible to restore unitarity
of the scattering amplitudes in the resonance region by including the
Higgs width in the Higgs propagator, which amounts to a modification
of Feynman's expansion of the scattering operator by incorporating
certain higher-order contributions.  But we shall employ the more general
K-matrix and Pad\'e methods. It is
interesting that upon unitarization the effect of an open
$HH$ channel can be felt even in the processes $W^+W^-\rightarrow
W^+W^-$, $W^+W^-\rightarrow ZZ$ and $ZZ\rightarrow ZZ$. Unitarization
is also necessary to properly study the breakdown of the perturbative
calculations at high energies for various values of the Higgs mass
because unitarization has a damping effect on the scattering amplitudes
and shifts the breakdown of the perturbative treatment to higher energies.

	In addition to deriving the unitarized amplitudes for the $W$,
$Z$ and Higgs scattering, we shall also compute the cross
sections for $W^+W^-$ and $ZZ$ production in $pp$ collisions leading to
the detection of the Higgs boson.  The mass of the Higgs boson remains
unknown.  However, since the Higgs doublet is introduced in the
Standard Model to generate the masses of the gauge bosons, it is
generally accepted that the Higgs boson mass could not exceed the $W$
and $Z$ masses by a large factor.  Therefore, a search for the Higgs
boson of mass up to 1~TeV should enable us to discover this particle,
if it exists.

\section{INTERACTION OF GAUGE AND HIGGS BOSONS}
According to the Goldstone-boson equivalence theorem,
which is applicable to scattering processes in the region $s,m_H^2 \gg m_W^2$,
the Lagrangian density for a system of $w^\pm=W_L^\pm$, $z=Z_L$ and $H$ bosons
is expressible as
\begin{equation}
L=L_0+L_{int},
\end{equation}
where $L_0$ represents the free Lagrangian density, and
\begin{eqnarray}
L_{int}&=&-\lambda(w^+w^- + {\textstyle\frac{1}{2}}z^2
	+{\textstyle\frac{1}{2}}H^2)^2\nonumber\\
& &-\kappa(w^+w^- + {\textstyle\frac{1}{2}}z^2+{\textstyle\frac{1}{2}}H^2)H,
\end{eqnarray}
with
\begin{equation}
\lambda=g^2\frac{m_H^2}{8m_W^2}\ ,\qquad
	\kappa=g\frac{m_H^2}{2m_W}=\sqrt{2\lambda}\,m_H.
\end{equation}

	The elimination of divergences requires mass renormalizations, which
can be carried out by the procedure followed by Marciano and
Willenbrock\cite{MW}.
The counterterms generated by the renormalizations
$m_W^2 \rightarrow m_W^2-\delta m_W^2$
and $m_H^2\rightarrow m_H^2-\delta m_H^2$ are
\begin{eqnarray}
\delta L &=& \delta m_W^2(w^+w^-+{\textstyle\frac{1}{2}}z^2)
	+{\textstyle\frac{1}{2}}\delta m_H^2H^2\nonumber\\
& & + \delta\lambda(w^+w^-+{\textstyle\frac{1}{2}}z^2
	+{\textstyle\frac{1}{2}}H^2)^2\nonumber\\
& & + \delta\kappa(w^+w^-+{\textstyle\frac{1}{2}}z^2
	+{\textstyle\frac{1}{2}}H^2)H,
\end{eqnarray}
where the first two terms arise from $L_0$, and
\begin{equation}
\frac{\delta\lambda}{\lambda}=\frac{\delta m_H^2}{m_H^2}\ -\
	\frac{\delta m_W^2}{m_W^2}\ ,\qquad
\frac{\delta\kappa}{\kappa}=\frac{\delta m_H^2}{m_H^2}\ -\
	\frac{\delta m_W^2}{2m_W^2}\ .
\end{equation}

	It should be noted that while the contributions of the $w^\pm$
and $z$ tadpoles vanish according to dimensional
regularization, the contributions of the Higgs tadpoles are
nonvanishing.  These nonvanishing contributions can be absorbed in
$\delta m_W^2$ and $\delta m_H^2$.  But, it is simpler to follow
Taylor's prescription\cite{taylor}, which amounts to introducing the
additional counterterms
\begin{equation}
\delta L^\prime = -\delta\eta H - \delta\eta\frac{g}{2m_W}(w^+w^-
	+\frac{1}{2}z^2
	+\frac{1}{2}H^2),\label{del.lag}
\end{equation}
and choosing $\delta\eta$ such that the contribution from the first term in
(\ref{del.lag}) cancels that from the Higgs tadpole-tree diagrams, while the
contributions from the second term are included in $\delta m_W^2$ and
$\delta m_H^2$.

	The renormalization constants $\delta m_W^2$, $\delta m_H^2$, and
$\delta\eta$ are found to be
\begin{eqnarray}
\frac{\delta m_W^2}{m_W^2}&=&\frac{\lambda}{16\pi^2}\ ,\nonumber\\
\frac{\delta m_H^2}{m_H^2}&=&-\frac{3\lambda}{16\pi^2}\left(
	\frac{4}{\hat{\epsilon}}
	+8-\sqrt{3}\pi\right),\\
\frac{\delta\eta}{m_H^3}&=&-\frac{3}{2}\sqrt{2\lambda}
	\left(\frac{1}{\hat{\epsilon}}+1\right),\nonumber
\end{eqnarray}
where
\begin{equation}
\frac{1}{\hat{\epsilon}}=\frac{2}{4-D}-\gamma_E
	+\ln\left(\frac{4\pi \mu^2}{m_H^2}\right).
\end{equation}

\section{$W$, $Z$ AND HIGGS SCATTERING}
\label{scattering}
	The treatment of the Feynman diagrams for the $w^\pm$, $z$ and
$H$ scattering to one-loop requires the evaluation of
complicated loop integrals.  Our results for these integrals are
summarized in the Appendix, where $A$, $B_n$, $C_n$ and $D_n$ refer to
the integrals arising from tadpole, bubble, triangle and box diagrams,
respectively.  For performing calculations, we have found it convenient
to take $m_H$ as the unit of mass and consequently of energy as well.
Thus, in addition to setting $c=\hbar=1$, we have also set $m_H=1$.

	The contributions of some of the scattering diagrams can be
expressed in terms of the two- and three-point functions.  The
two-point functions corresponding to the diagrams in
Fig.~\ref{two.point} are
\widetext
\begin{eqnarray}
\Pi_w(s)&=&-2\frac{\lambda}{16\pi^2}\left[B_1(s)-1\right],\nonumber\\
\\
\Pi_H(s)&=&-\frac{\lambda}{16\pi^2}\left[3B_2(s)+9B_3(s)+3A(1)\nonumber
	-27+3\pi\sqrt{3}\,\right],
\end{eqnarray}
which also yield the external wave-function renormalization factors
\begin{eqnarray}
Z_w&=&1-\frac{\lambda}{16\pi^2}\ ,\nonumber\\
\\
Z_H&=&1+\frac{\lambda}{16\pi^2}\left(12-2\pi\sqrt{3}\,\right).\nonumber
\end{eqnarray}
The three-point functions, which correspond to Figs.~\ref{three.point}(a),
\ref{three.point}(b) and \ref{three.point}(c) are
\begin{eqnarray}
\Gamma_a(s)&=&\frac{\lambda^{3/2}}{16\pi^2\sqrt{2}}
	\Bigl[2\cdot4B_1(0)+10B_2(s)+6B_3(s)\nonumber\\
	& &\hspace{1.in}\left. +4C_1(s)+12C_2(s)-49+6\pi\sqrt{3}\,
	\right],\nonumber\\
\Gamma_b(s)&=&\frac{\lambda^{3/2}}{16\pi^2\sqrt{2}}
	\Bigl[4B_1(0)+10B_2(1)+6B_3(1)+4B_1(s)\nonumber\\
	& &\hspace{1.in}\left.+12C_3(s)+4C_4(s)-49+6\pi\sqrt{3}\,\right],\\
\Gamma_c(s)&=&3\frac{\lambda^{3/2}}{16\pi^2\sqrt{2}}
	\Bigl[2\cdot2B_2(1)+2\cdot6B_3(1)+2B_2(s)+6B_3(s)\nonumber\\
	& &\hspace{1.in}\left.+ 4C_5(s)+36C_6(s)
	-49+6\pi\sqrt{3}\,\right].\nonumber
\end{eqnarray}

	The Feynman diagrams for the processes $w^+w^-\rightarrow zz$,
$w^+w^-\rightarrow HH$ and
$HH \rightarrow HH$ are shown in Figs.~\ref{wwzz}, \ref{wwhh} and
\ref{hhhh}, and the scattering amplitudes
corresponding to these diagrams can now be written as\cite{wznote}
\begin{eqnarray}
M(w^+w^-\leftrightarrow zz)&=&M(s,t,u),\nonumber\\
M(w^+w^-\leftrightarrow HH)&=&M'(s,t,u),\\
M(HH\rightarrow HH)&=&M''(s,t,u),\nonumber
\end{eqnarray}
where
\begin{eqnarray}
{M}(s,t,u)
&=&-2\lambda\left[1+\frac{1}{s-1}\right]Z_w^2
\ -2\lambda\frac{\Pi_H(s)}{(s-1)^2}
\ +2\cdot 2\frac{\lambda^{1/2}}{\sqrt{2}}\frac{\Gamma_a(s)}{s-1}\nonumber\\
& &+\frac{\lambda^2}{16\pi^2}\left[14B_2(s)+2B_3(s)+4B_2(t)
	+4B_2(u)\right]\nonumber\\
& &+2\cdot4\frac{\lambda^2}{16\pi^2}\left[C_2(s)+C_1(s)+C_1(t)+C_1(u)
	\right]\nonumber\\
& &+4\frac{\lambda^2}{16\pi^2}\left[D_1(s,t)+D_1(s,u)\right]
	-2\frac{\lambda^2}{16\pi^2}\left(25-3\pi\sqrt{3}\,\right),\label{wz}
\end{eqnarray}
\begin{eqnarray}
{M'}(s,t,u)&=&
-2\lambda\left[1+\frac{3}{s-1}+\frac{1}{t}+\frac{1}{u}\right]Z_wZ_H\nonumber\\
& &-\lambda\left[6\frac{\Pi_H(s)}{(s-1)^2}
	+2\frac{\Pi_w(t)}{t^2} +2\frac{\Pi_w(u)}{u^2}\right]\nonumber\\
& &+\frac{\lambda^{1/2}}{\sqrt{2}}\left[6\frac{\Gamma_a(s)}{s-1}
	+2\frac{\Gamma_c(s)}{s-1} +2\cdot 2\frac{\Gamma_b(t)}{t}
	+2\cdot 2\frac{\Gamma_b(u)}{u}\right]\nonumber\\
& &+\frac{\lambda^2}{16\pi^2}\left[10B_2(s)+6B_3(s)+4B_1(t)+4B_1(u)
	\right]\nonumber\\
& &+\frac{\lambda^2}{ 16\pi^2}\left[4C_1(s)+12C_2(s)
	+2\cdot4C_3(t)+2\cdot4C_3(u)
	\right.\nonumber\\
	& & \qquad \left.+2\cdot12C_4(t)+2\cdot12C_4(u)
	+20C_5(s)+36C_6(s)\right]\nonumber\\
& &+\frac{\lambda^2}{ 16\pi^2}\left[4D_2(s,t)+4D_2(s,u)+12D_3(s,t)
	+12D_3(s,u)\right.\nonumber\\
	& & \left.\qquad +36D_4(s,t)+36D_4(s,u)\right]
	-2\frac{\lambda^2}{16\pi^2}\left(25-3\pi\sqrt{3}\,\right),
\end{eqnarray}
\begin{eqnarray}
{M''}(s,t,u)&=&
-6\lambda\left[1+\frac{3}{s-1} +\frac{3}{t-1}
	+\frac{3}{u-1}\right]Z_H^2\nonumber\\
& &-\lambda\left[16\frac{\Pi_H(s)}{(s-1)^2}
	+18\frac{\Pi_H(t)}{(t-1)^2}
	+18\frac{\Pi_H(u)}{(u-1)^2}\right]\nonumber\\
& &+2\cdot 6\frac{\lambda^{1/2}}{\sqrt{2}}\left[\frac{\Gamma_c(s)}{s-1}
	+\frac{\Gamma_c(t)}{t-1}+\frac{\Gamma_c(u)}{u-1}\right]\nonumber\\
& &+\frac{\lambda^2}{16\pi^2}\left\{6\left[B_2(s)+B_2(t)+B_2(u)\right]
	+18\left[B_3(s)+B_3(t)+B_3(u)\right]\right\}\nonumber\\
& &+\frac{\lambda^2}{16\pi^2}\left\{2\cdot 12\left[C_5(s)+C_5(t)+C_5(u)\right]
	+2\cdot 108\left[C_6(s)+C_6(t)+C_6(u)\right]\right\}\nonumber\\
& &+\frac{\lambda^2}{16\pi^2}\left\{
	3\cdot4\left[D_5(s,t)+D_5(s,u)+D_5(u,t)\right]\right.\nonumber\\
& &\left.+324\left[D_6(s,t)+D_6(s,u)+D_6(u,t)\right]\right\}
	-6\frac{\lambda^2}{16\pi^2}\left(25-3\pi\sqrt{3}\,\right).
\end{eqnarray}
We have also included the external wave-function
renormalization factors, and it is understood that only terms
to order $\lambda$ are
to be retained in $Z_w^2$, $Z_wZ_H$, and $Z_H^2$.

	It follows from an SO(3) symmetry associated with the gauge boson
interactions that
\begin{eqnarray}
M(w^+w^-\rightarrow w^+w^-)&=&M(s,t,u)+M(t,s,u),\nonumber\\
M(zz\rightarrow zz)&=&M(s,t,u)+M(t,s,u)+M(u,t,s),\\
M(zz\leftrightarrow HH)&=&M'(s,t,u).\nonumber
\end{eqnarray}
It should also be noted that, according to the notation used in the Appendix,
$p=({\bf p},ip_0)$, and the Mandelstam variables appearing in the scattering
amplitudes are
\begin{equation}
s=-(p_1+p_2)^2, \qquad t=-(p_1-p_3)^2,\qquad u=-(p_1-p_4)^2.
\end{equation}
\narrowtext

	The S-wave projection of the amplitude $M(s,t,u)$ is given by
\cite{pnote}
\begin{equation}
a(s)=\frac{1}{32\pi}\left(\frac{4 |{\bf p}_f| |{\bf p}_i|}{s}\right)^{1/2}
	\int_{-1}^1 d\!\cos\theta \:M(s,t,u), \label{pwform}
\end{equation}
and let $b(s)$, $c(s)$ and $d(s)$ denote similar projections of $M(t,s,u)$,
$M'(s,t,u)$ and $M''(s,t,u)$, respectively.  We can then express the
S-wave amplitude
for the coupled $w^+w^-$, $zz$ and $HH$ channels as
\begin{equation}
a_0=\left( \renewcommand{\arraystretch}{1.5}
\begin{array}{ccc}
a+b&\displaystyle\frac{a}{\sqrt{2}}&\displaystyle\frac{c}{\sqrt{2}}\\
\displaystyle\frac{a}{\sqrt{2}}&\displaystyle\frac{a}{2}
	+b&\displaystyle\frac{c}{2}\\
\displaystyle\frac{c}{\sqrt{2}}&\displaystyle\frac{c}{2}&\displaystyle
	\frac{d}{2}
\end{array} \right)\ ,\label{pwamp}
\end{equation}
where we have also introduced appropriate factors to eliminate
the identical-particle
restrictions on the $zz$ and $HH$ phase spaces.

\section{UNITARIZED SCATTERING AMPLITUDES AND CROSS SECTIONS}
	Since Feynman's expansion of the scattering operator is not
unitary order by order, the amplitudes obtained in the
preceding section violate unitarity, and we shall apply two
unitarization schemes as mentioned earlier.

	In the K-matrix formalism\cite{gupta}, the S~matrix for the
gauge and Higgs boson scattering is expressed as
\begin{equation}
S=\frac{1-\frac{1}{2}iK}{1+\frac{1}{2}iK}
\end{equation}
with the expansion
\begin{equation}
K=K_1+K_2+\ldots,
\end{equation}
where $K_n$ is of $n$th order in the coupling parameter $\lambda$. It
follows from the unitarity of S that K is Hermitian, and consequently
the K-matrix expansion ensures unitarity order by order.  The $K_n$ are
related to the $S_n$ as
\widetext
\begin{eqnarray}
K_n&=&iS_n+i\left({\textstyle -\frac{1}{2}}\right)\sum_{n_1
+n_2=n} S_{n_1}S_{n_2}
	+\cdots\nonumber\\
& &+i\left({\textstyle -\frac{1}{2}}\right)^{p-1}\sum_{n_1+n_2+\cdots+n_p=n}
	S_{n_1}S_{n_2}\cdots S_{n_p}+\cdots,\label{kmatrix}
\end{eqnarray}
\narrowtext
and in particular,
\begin{equation}
K_1=iS_1,\qquad K_2=iS_2-\frac{1}{2}iS_1 S_1,
\end{equation}
which leads to
\begin{equation}
K_2={\rm Re\,} ( iS_2).
\end{equation}
The relation (\ref{kmatrix}) shows that the $K_n$ can be given a simple
physical interpretation: The first term on the right side of
(\ref{kmatrix}) contains the contributions of all possible nth order
transitions, while the remaining terms cancel the contributions of
those transitions which are made up of two or more real transitions.

	Let
\begin{equation}
a_J=a^{(1)}_J+a^{(2)}_J \label{feynman}
\end{equation}
be the Feynman amplitude to one loop for the $J$th partial wave in a
one-channel
scattering process. Then the K-matrix amplitude is
\begin{equation}
a^K_J=\frac{a^{(1)}_J+{\rm Re\,} a^{(2)}_J}{1-i\left(a^{(1)}_J+{\rm Re\,}
a^{(2)}_J\right)}\ ,
\end{equation}
which satisfies the unitarity relation
\begin{equation}
{\rm Im\,} a^K_J = |a^K_J|^2. \label{kunit}
\end{equation}

	The Pad\'e method\cite{pade} focuses on the mathematics of a
series rather than the physics, and it approximates a series in
the form of a ratio of two polynomials. For (\ref{feynman}), the Pad\'e
approximant is given by
\begin{equation}
a^P_J=\frac{{a^{(1)}_J}^2}{a^{(1)}_J-a^{(2)}_J}\ .\label{pade}
\end{equation}
We can also write (\ref{pade}) as
\[
a^P_J=\frac{{a^{(1)}_J}^2}{a^{(1)}_J-{\rm Re\,}
	a^{(2)}_J-i{\rm Im\,} a^{(2)}_J}\ ,
\]
which gives
\[
{\rm Im\,} a^P_J=|a^P_J|^2\frac{{\rm Im\,} a^{(2)}_J}{{a^{(1)}_J}^2}\ ,
\]
and to ensure that the Pad\'e approximant satisfies the unitarity
condition
\begin{equation}
{\rm Im\,} a^P_J = |a^P_J|^2, \label{punit}
\end{equation}
we must have
\begin{equation}
{\rm Im\,} a^{(2)}_J={a^{(1)}_J}^2\ .\label{unit}
\end{equation}

	For multichannel scattering, $a_J$ is expressible as a matrix,
and in order to satisfy (\ref{kunit}) and (\ref{punit}) the
K-matrix and Pad\'e amplitudes take the form
\widetext
\begin{eqnarray}
a^K_J&=&\left(a^{(1)}_J+{\rm Re\,} a^{(2)}_J\right)
	\left[1-i\left(a^{(1)}_J+{\rm Re\,} a^{(2)}_J\right)\right]^{-1},\\
a^P_J&=&a^{(1)}_J\left(a^{(1)}_J-a^{(2)}_J\right)^{-1}a^{(1)}_J,
\end{eqnarray}
\narrowtext
which we have used to obtain the unitarized S-wave amplitudes for the coupled
$w^+w^-$, $zz$ and $HH$ channels from (\ref{pwamp}).

	In Figs.~\ref{amps1} and~\ref{amps2}, we have plotted absolute
values of the S-wave amplitudes for scatterings among the
$w^+w^-$, $zz$ and $HH$ channels as functions of the center-of-mass
energy $\sqrt s$ for $m_H=500$~GeV and 1000~GeV.  The figures show
the Feynman amplitudes as well as the unitarized K-matrix and
Pad\'e amplitudes for various processes of physical interest.  The
unitarity violation of the Feynman amplitude is more severe
for the larger value of the Higgs mass, and in Fig.~\ref{amps2}(a) the
amplitude does not return below unity after reaching the
resonance peak.  The K-matrix and Pad\'e amplitudes agree at lower energies,
but the Pad\'e amplitudes seem to exhibit an interesting but odd behavior
at higher energies, whose physical significance is unclear\cite{newpade}.

	For the detection of the Higgs boson through gauge boson
scattering in $pp$~collisions, we have used the unitarized
amplitudes to compute the invariant-mass distributions for the
processes
\[
 pp\rightarrow W_L^+W_L^-\ {\rm or}\ Z_LZ_L\rightarrow W^+_LW^-_LX
\]
and
\[
 pp\rightarrow W_L^+W_L^-\ {\rm or}\ Z_LZ_L\rightarrow Z_LZ_LX
\]
at the SSC energy of $\sqrt s=40$~TeV for $m_H=500$~GeV and 1000~GeV by
means of the \hbox{effective-W} approximation\cite{repko}.  The
distributions corresponding to the K-matrix and Pad\'e expansions are
shown in Fig.~\ref{cross}.

	Finally, in Fig.~\ref{compare} we have compared the K-matrix
$W_L^+W_L^- \rightarrow W_L^+W_L^-$ amplitudes
without and with the one-loop correction.  We have not provided a similar
comparison with the Pad\'e
method, which is unable to unitarize the tree amplitudes.
The one-loop correction in Fig.~\ref{compare} is more pronounced for
$m_H=1000$~GeV, and it increases for larger values of $\sqrt s$, but amounts
to only $10\%$ at $\sqrt s =1500$~GeV.  We also observe that, according
to Fig.~\ref{cross}(c), the $W_L^+W_L^-$ pair production
cross section in $pp$ collisions reaches its peak value at
$m_{WW}=850$~GeV, and falls to one-fifteenth of  this value at
$m_{WW}=1500$~GeV.
Thus, Figs.~\ref{cross} and~\ref{compare} show that the bulk of the pair
production occurs at those values of $m_{WW}$ for which the one-loop correction
to the $W_L^+W_L^- \rightarrow W_L^+W_L^-$ amplitude is significant but
not unacceptably large.  Altogether, a perturbative treatment based on
the K-matrix appears to yield sensible results for the gauge-boson
and Higgs scattering at SSC energies.

\acknowledgments
	This work was supported in part by the U.S. Department of Energy
under Grant No. DE-FG02-85ER40209 and the National Science Foundation
under Grant No. PHY-90-06117.

\cleardoublepage
\widetext
\appendix
\section*{EVALUATION OF LOOP INTEGRALS}

	The results for all the loop integrals encountered in our
	treatment are summarized in this appendix.  Many of these
integrals do not appear in the absence of the $HH$ channel, and they
have not been evaluated by earlier authors\cite{ampnote}.  As explained
in Sec.~\ref{scattering}, we have set $m_H=1$, and our space-time
metric is such that $p=({\bf p},ip_0)$. Our notation for the loop
integrals is similar but somewhat different from the Passarino-Veltman
notation \cite{PV}.
\subsection{Tadpole integral}
\begin{eqnarray*}
{\cal A}(m)&=&\int \frac{{\cal D}^Dq}{(2\pi)^D}
	\frac{1}{[q^2+m^2-i\varepsilon]}\\
  &=&-\frac{im^2}{16\pi^2}\left[\frac{1}{\hat\epsilon}+A(m^2)\right],\\
A(1)&=&1.
\end{eqnarray*}

\subsection{Bubble integrals }
\sloppy
\begin{eqnarray*}
& &s=-p^2,\qquad \beta=\sqrt{1-4/s},\\
{\cal B}(p,m_1,m_2)&=&\int \frac{{\cal D}^Dq}{(2\pi)^D}
	\frac{1}{[q^2+m_1^2-i\varepsilon]
	[(p-q)^2+m_2^2-i\varepsilon]}\\
 &=&\frac{i}{16\pi^2}\left[\frac{1}{\hat\epsilon}+B_n(s)\right].
\end{eqnarray*}
\leftline{\it Bubble 1. $m_1=1$, $m_2=0$:}
\begin{eqnarray*}
B_1(s)&=& 2 - \frac{s-1}{s}\ln |1-s| +\theta(s-1)i\pi \frac{s-1}{s}
	\mbox{\quad for all $s$,}\\
&=&1 \mbox{\quad for $s=0$,}\\
&=&2 \mbox{\quad for $s=1$.}
\end{eqnarray*}
\leftline{\it Bubble 2. $m_1=m_2=0$: }
\begin{eqnarray*}
B_2(s)&=& 2 - \ln |s| +\theta(s)i\pi \mbox{\quad for all $s$,} \\
&=&2+i\pi \mbox{\quad for $s=1$.}
\end{eqnarray*}
\leftline{\it Bubble 3. $m_1=m_2=1$:}
\begin{eqnarray*}
B_3(s)&=& 2 -\beta\ln\left|\frac{1+\beta}{1-\beta}\right| +\theta(s-4)i\pi\beta
	\mbox{\quad for $s<0$ and $s>4$,}\\
&=& 2-2\sqrt{\frac{4}{s}-1}\ \arcsin\frac{\sqrt{s}}{2}
	\mbox{\quad for $0<s<4$,} \\
&=&0 \mbox{\quad for $s=0$,}\\
&=&2-\frac{2\pi}{\sqrt{3}} \mbox{\quad for $s=1$.}
\end{eqnarray*}

\subsection{Triangle integrals{\rm \protect\cite{spence}}}
\sloppy
\[ s=-(p_1+p_2)^2,\quad
\beta=\sqrt{1-4/s}\ ,\quad
s_\pm =\frac{1\pm\beta}{2}\ ,\quad
\gamma_\pm= e^{\pm i\pi/3}\ , \]
\begin{eqnarray*}
{\cal C}(p_1,p_2,m_1,m_2,m_3)&=&\int \frac{{\cal D}^Dq}{(2\pi)^D}
	\frac{1}{[q^2+m_1^2-i\varepsilon] [(q-p_1)^2+m_2^2-i\varepsilon]
	[(q-p_1-p_2)^2+m_3^2-i\varepsilon]}\\
 &=&-\frac{i}{16\pi^2}C_n(s).
\end{eqnarray*}
\leftline{\it Triangle 1. $m_1=m_3=0$, $m_2=1$, $p_1^2=p_2^2=0$:}
\begin{eqnarray*}
C_1(s)&=&\frac{1}{s}\left[\frac{\pi^2}{6} - {\rm Sp}(s+1)-
	\theta(s)i\pi \ln(s+1)\right] \mbox{\quad for all $s$.}
\end{eqnarray*}
\leftline{\it Triangle 2. $m_1=m_3=1$, $m_2=0$, $p_1^2=p_2^2=0$:}
\begin{eqnarray*}
C_2(s)&=&-\frac{2}{s}\left[ {\rm Sp}(ss_+)+{\rm Sp}(ss_-)
	+\theta(s-4)2i\pi\ln(\sqrt{s}s_+) \right] \mbox{\quad for all $s$,}\\
&=&-\frac{4}{s}\left(\arcsin\frac{\sqrt{s}}{2}\right)^2
	\mbox{\quad for $0<s<4$.}
\end{eqnarray*}
\leftline{\it Triangle 3. $m_1=1$, $m_2=m_3=0$, $p_1^2=0$, $p_2^2=-1$:}
\begin{eqnarray*}
C_3(s)&=&\frac{-1}{s-1}\left[\frac{\pi^2}{6}+{\rm Sp}(s(2-s))-{\rm Sp}(s)
	-{\rm Sp}(2-s)\right.\\
& &\hspace{0.5in}+\theta(s-1)i\pi\ln(s) \Biggr]
	\mbox{\quad for all $s$.}
\end{eqnarray*}
\leftline{\it Triangle 4. $m_1=0$, $m_2=m_3=1$, $p_1^2=0$, $p_2^2=-1$:}
\begin{eqnarray*}
C_4(s)&=&\frac{1}{s-1} \left[ \frac{\pi^2}{6}-{\rm Sp}(s)
	-{\rm Sp}\left(\frac{s^2}{s^2-s+1}\right)+{\rm
Sp}\left(\frac{s}{s^2-s+1}\right)
	-2{\rm Sp}\left(\frac{\gamma_+}{s-1+\gamma_+}\right)\right.\\
& & \left.\hspace{0.5in}+2{\rm Sp}\left(\frac{s\gamma_+}{s-1+\gamma_+}\right)
	+\theta(s-1)i\pi\ln\left(\frac{s}{s^2-s+1}\right) \right]
	\mbox{\quad for all $s$.}
\end{eqnarray*}
\leftline{\it Triangle 5. $m_1=m_2=m_3=0$, $p_1^2=p_2^2=-1$:}
\begin{eqnarray*}
C_5(s)&=&-\frac{2}{s}{\rm Re\,}\left\{ \frac{1}{\beta}\left[ {\rm Li_2}(1-ss_+)
  	-{\rm Li_2}(1-ss_-)\right]\right\}-\theta(s)\frac{4i\pi}{s\beta}
	\ln(\sqrt{s}s_+)
	\mbox{\quad for all $s$.}
\end{eqnarray*}
\leftline{\it Triangle 6. $m_1=m_2=m_3=1$, $p_1^2=p_2^2=-1$:}
\begin{eqnarray*}
C_6(s)&=&\frac{1}{s} {\rm Re\,}\left\{\frac{1}{\beta}\left[
	{\rm Li_2}(ss_--1)-{\rm Li_2}(ss_+-1)
	+{\rm Li_2}\left(\frac{ss_--1}{s-3}\right)
	-{\rm Li_2}\left(\frac{ss_+-1}{s-3}\right) \right. \right. \\[3pt]
 & & \left. \left. \hspace*{0.25in}
	+2{\rm Li_2}\left(\frac{s_+}{s_+-\beta\gamma_+}\right)
	-2{\rm Li_2}\left(\frac{s_-}{s_+-\beta\gamma_+}\right)
	+2{\rm Li_2}\left(\frac{s_+}{s_-+\beta\gamma_+}\right)
	-2{\rm Li_2}\left(\frac{s_-}{s_-+\beta\gamma_+}\right)
	\right]\right\} \\[6pt]
 & & \hspace*{0.5in}-\theta(s-4)\frac{i\pi}{s\beta}\ln(s-3)
	\mbox{\quad for all $s$.}
\end{eqnarray*}

\subsection{Box integrals}
\sloppy
\begin{eqnarray*}
s&=&-(p_1+p_2)^2,\quad t=-(p_3-p_1)^2,\\
s_\pm &=&\frac{1\pm\beta}{2}\ ,\quad \beta=\sqrt{1-4/s}\ ,
	\quad t_+=\frac{1+\sqrt{1-4/t}}{2}\ ,
\quad\gamma_\pm=e^{\pm i\pi/3}\ ,
\end{eqnarray*}
\begin{eqnarray*}
& &{\cal D}(p_1,p_2,p_3,m_1,m_2,m_3,m_4)\\
&=&\int \!\!\frac{{\cal D}^Dq}{(2\pi)^D}
\frac{1}{[q^2+m_1^2-i\varepsilon] [(q-p_1)^2+m_2^2-i\varepsilon]
[(q-p_1-p_2)^2+m_3^2-i\varepsilon][(q-p_1-p_2+p_4)^2+m_4^2-i\varepsilon]}\\
&=&\frac{i}{16\pi^2}D_n(s,t).
\end{eqnarray*}

\leftline{\it Box 1. \rm  $m_1=m_3=1$, $m_2=m_4=0$,
	$p_1^2=p_2^2=p_3^2=p_4^2=0$:}
\begin{eqnarray*}
D_1(s,t)&=&\frac{1}{st\lambda} \left[
{\rm Sp}\left(\frac{\lambda_+ -1}{\lambda_+ -s_+}\right)
+{\rm Sp}\left(\frac{\lambda_+ -1}{\lambda_+ -s_-}\right)
+{\rm Sp}\left(\frac{(t+1)\left(\lambda_+ -1\right)}{\lambda_+(t+1) -1}\right)
+{\rm Sp}\left(\frac{\lambda_+}{\lambda_+ -1}\right)\right. \\
& &\qquad \left.
-{\rm Sp}\left(\frac{(t+1) \lambda_+}{\lambda_+ (t+1) -1}\right)
-{\rm Sp}\left(\frac{\lambda_+}{\lambda_+ -s_+}\right)
-{\rm Sp}\left(\frac{\lambda_+}{\lambda_+ -s_-}\right) \right]
	-\left[ \lambda_+ \rightarrow \lambda_- \right] \\
& &\qquad + \theta(s-4)\frac{i\pi}{st\lambda}\ln
	\frac{ (s_+ -\lambda_+)(s_- -\lambda_-)}{(s_- -\lambda_+)(s_+
 	-\lambda_-)}\
+\theta(t)\frac{i\pi}{st\lambda}\ln\frac{(1-\lambda)st-2t-2}{(1+\lambda)st
	-2t-2}\ , \\[12pt]
\lambda_\pm &=& \frac{  t+2 \pm t\lambda}{ 2(t+1) }\ ,\qquad
	\lambda\ =\ \sqrt{1-\frac{4}{s}\left(1+\frac{1}{t}\right) }\ .
\end{eqnarray*}
\vspace{0.5in}

\leftline{\it Box 2. \rm  $m_1=m_3=m_4=0$, $m_2=1$,
	$p_1^2=p_2^2=0$, $p_3^2=p_4^2=-1$:}
\begin{eqnarray*}
D_2(s,t)&=&\frac{1}{\lambda} \left[
{\rm Sp}  \left(\frac{(s-1) \left(\lambda_+ -1 \right) }{
	\lambda_+ (s-1) +1} \right)
-{\rm Sp} \left(\frac{(s-1) \lambda_+}{ \lambda_+(s-1) +1} \right)
+{\rm Sp} \left(\frac{\left(\lambda_+ -1 \right) t}{  t\lambda_+
	-t +1} \right) \right.\\
& & \left. -{\rm Sp} \left(\frac{t \lambda_+}{ t\lambda_+ -t +1} \right)
-{\rm Sp} \left(\frac{2 \left(\lambda_+ -1 \right)}{ 2 \lambda_+ -1} \right)
+{\rm Sp} \left(\frac{2 \lambda_+}{ 2 \lambda_+ -1} \right)
+{\rm Sp} \left(\frac{\lambda_+}{ \lambda_+ -1} \right) \right]\\
& &-\left[ \lambda_+ \rightarrow \lambda_- \right]

+\theta(s)\frac{i\pi}{\lambda}\ln\frac{\left[(t-1)(2+st)-t\lambda\right](1+s)}{
	(t-1)(2+st+s^2-2s)+4s+(t+s-2)\lambda}\ ,\\[12pt]
\lambda_\pm&=&\frac{(t-1)s -2t+4 \pm \lambda}{2\left(st-t+2\right)}\ ,\qquad
\lambda=\sqrt{s} \sqrt{ (t-1)^2 s +4 t -8}\ .
\end{eqnarray*}

\leftline{\it Box 3. \rm  $m_1=m_4=1$, $m_2=m_3=0$,
	$p_1^2=p_4^2=0$, $p_2^2=p_3^2=-1$:}
\begin{eqnarray*}
D_3(s,t)&=&\frac{1}{\lambda} \left[
{\rm Sp} \left(\frac{\lambda_+ -1}{ \lambda_+ -\gamma_+}\right)
+{\rm Sp} \left(\frac{\lambda_+ -1}{ \lambda_+ -\gamma_-} \right)
-{\rm Sp}\left(\frac{(s+t-2)\left(\lambda_+ -1\right)}{\lambda_+
	(s+t-2) +1}\right)
+{\rm Sp} \left(\frac{\lambda_+}{ \lambda_+ -1}\right)\right.\\
& &\qquad
+{\rm Sp} \left(\frac{\left(s +t -2 \right) \lambda_+}{\lambda_+(s+t-2)
	+1} \right)
-{\rm Sp} \left(\frac{\left(\lambda_+ -1 \right) t}{t \lambda_+ -1} \right)
+{\rm Sp} \left(\frac{t \lambda_+}{ t \lambda_+ -1} \right)
+{\rm Sp} \left(\frac{2\lambda_+ -2}{ 2\lambda_+ -1} \right) \\
& &\qquad \left.
-{\rm Sp} \left(\frac{2 \lambda_+}{ 2 \lambda_+ -1} \right)
-{\rm Sp} \left(\frac{\lambda_+}{ \lambda_+ -\gamma_+} \right)
-{\rm Sp} \left(\frac{\lambda_+}{ \lambda_+ -\gamma_-} \right) \right]
-\left[ \lambda_+ \rightarrow \lambda_- \right] ,\\[12pt]
\lambda_\pm&=&\frac{(t+1)s+t^2-2t+1 \pm\lambda}{2(st-2t+t^2+2)}\ ,\\[8pt]
\lambda&=&\sqrt{s^2t^2-2s^2t+s^2+2st^3-6st^2+6st-6s+t^4-4t^3+6t^2-4t+1}\ .
\end{eqnarray*}

\vspace{0.25in}
\leftline{\it Box 4. \rm  $m_1=m_3=m_4=1$, $m_2=0$,
	$p_1^2=p_2^2=0$, $p_3^2=p_4^2=-1$:}
\begin{eqnarray*}
D_4(s,t)&=&\frac{1}{\lambda}\left[
{\rm Sp} \left(\frac{y_{1a}-1}{y_{1a}-\rho_{a+}} \right)
-{\rm Sp} \left(\frac{y_{2a}-1}{y_{2a}-\rho_{a+}} \right)
-{\rm Sp} \left(\frac{y_{1a}}{y_{1a}-\rho_{a+}} \right)
+{\rm Sp} \left(\frac{y_{2a}}{y_{2a}-\rho_{a+}} \right) \right. \\
& &\qquad -{\rm Sp} \left(\frac{y_{1b}-1}{y_{1b}-\rho_{b+}} \right)
 +{\rm Sp} \left(\frac{y_{2b}-1}{y_{2b}-\rho_{b+}} \right)
+{\rm Sp} \left(\frac{y_{1b}}{y_{1b}-\rho_{b+}} \right)
 -{\rm Sp} \left(\frac{y_{2b}}{y_{2b}-\rho_{b+}} \right) \\
& & \qquad \left. -{\rm Sp}\left(\frac{y_{3b}-1}{y_{3b}-s_+}\right)
+{\rm Sp}\left(\frac{y_{3b}}{y_{3b}-s_+}\right)
 +{\rm Sp}\left(\frac{y_{3a}-1}{y_{3a}-s_+}\right)
-{\rm Sp}\left(\frac{y_{3a}}{y_{3a}-s_+}\right) \right]\\
& &\qquad+\left[\rho_{a+},\rho_{b+},s_+\rightarrow\rho_{a-},\rho_{b-}
	,s_-\right]
-\theta (s-4)\frac{i\pi}{\lambda}\ln \left( \frac{\lambda_+}{\lambda_-}\cdot
	\frac{\beta-\lambda_-}{\beta-\lambda_+} \right)\ ,\\[8pt]
\rho_{a+}&=&1\ ,\qquad \rho_{a-}=\frac{(t-1)^2}{t^2-t+1}\ ,\\
\rho_{b\pm}&=&\frac{\left(t \gamma_\pm-1\right)(t-1)}{t^2-t+1}\ ,\\
\eta &=& \frac{(s t-s+\lambda)(t-1)}{2 (t^2-t+1)}\ ,\qquad
	\lambda=\sqrt{s} \sqrt{ (t-1)^2 s-4 t^2+4t-4}\ ,\\
\lambda_\pm&=&\frac{s(t+1-2ts_-) \pm\lambda}{2s(ss_-+t-1)}\ ,\\
\eta_a&=&-\frac{\left[(t-1)^2 s-2 t^2+3
	t-2\right](st-s+\lambda)-2(t^2-t+1)(t-1)s}{2
	(t^2-t+1)\lambda}\ ,\\
y_{1a}&=&\eta_a+\eta,\qquad y_{2a}=\frac{\eta_a}{1-\eta}\ ,\qquad
	y_{3a}=-\frac{\eta_a}{\eta}\ ,\\
\eta_b&=&-\frac{\left[\left(st-s+\lambda\right)(st-s-t+2)
	-2(t^2-t+1)s\right](t-1)}{2
	(t^{2}-t+1)\lambda}\ ,\\
y_{1b}&=&\eta_b+\eta,\qquad y_{2b}=\frac{\eta_b}{1-\eta}\ ,\qquad
	y_{3b}=-\frac{\eta_b}{ \eta}\ .
\end{eqnarray*}

\leftline{\it Box 5. \rm  $m_1=m_2=m_3=m_4=0$,
	$p_1^2=p_2^2=p_3^2=p_4^2=-1$: }
\begin{eqnarray*}
D_5(s,t)&=&\frac{1}{\lambda} \left[
{\rm Sp}\left(\frac{\lambda_+ -1}{ \lambda_+} \right)
-{\rm Sp}\left(\frac{\left(s -1 \right) \left(\lambda_+ -1
	\right) }{ (s-1) \lambda_+ +1} \right)
+{\rm Sp} \left(\frac{\left(s -1 \right) \lambda_+}{ (s-1) \lambda_+
	 +1} \right) \right. \\
& &\left. -{\rm Sp}\left(\frac{(t-1)\left(\lambda_+ -1\right)
	}{\lambda_+(t-1)-t} \right)
+{\rm Sp} \left(\frac{\left(t -1 \right) \lambda_+}{ \lambda_+
	(t-1)-t} \right)
-{\rm Sp} \left(\frac{\lambda_+}{ \lambda_+ -1} \right) \right]
	-\left[ \lambda_+ \rightarrow \lambda_- \right] \\
& &+\theta(s)\frac{i\pi}{\lambda}\ln\frac{\left[\lambda_+ (t-1)-t
	\right]\lambda_-}{\left[\lambda_-(t-1) -t\right]\lambda_+}\ ,\\[8pt]
\lambda_\pm&=&\frac{2t-st \pm\lambda}{2(t+s-ts) }\ ,\qquad
\lambda=st\sqrt{1 -\frac{4}{s t}}\ .
\end{eqnarray*}

\leftline{\it Box 6. \rm  $m_1=m_2=m_3=m_4=1$,
	$p_1^2=p_2^2=p_3^2=p_4^2=-1$:}
\begin{eqnarray*}
D_6(s,t)&=&\frac{\sqrt{1-4/t}}{\lambda}\left[
{\rm Sp}\left(\frac{y_{1a}-1}{y_{1a}-\rho_{a+}}\right)
-{\rm Sp}\left(\frac{y_{2a}-1}{y_{2a}-\rho_{a+}}\right)
-{\rm Sp}\left(\frac{y_{1a}}{y_{1a}-\rho_{a+}}\right)
+{\rm Sp}\left(\frac{y_{2a}}{y_{2a}-\rho_{a+}}\right) \right. \\
& & \qquad -{\rm Sp}\left(\frac{y_{1b}-1}{y_{1b}-\rho_{b+}}\right)
+{\rm Sp}\left(\frac{y_{2b}-1}{y_{2b}-\rho_{b+}}\right)
+{\rm Sp}\left(\frac{y_{1b}}{y_{1b}-\rho_{b+}}\right)
-{\rm Sp}\left(\frac{y_{2b}}{y_{2b}-\rho_{b+}}\right) \\
& & \qquad \left. +{\rm Sp}\left(\frac{y_{3a}-1}{y_{3a}-s_+}\right)
-{\rm Sp}\left(\frac{y_{3a}}{y_{3a}-s_+}\right)
-{\rm Sp}\left(\frac{y_{3b}-1}{y_{3b}-s_+}\right)
+{\rm Sp}\left(\frac{y_{3b}}{y_{3b}-s_+}\right) \right] \\
& &\qquad+\left[\rho_{a+},\rho_{b+},s_+\rightarrow\rho_{a-},\rho_{b-}
	,s_-\right]
 -\theta(s-4)\frac{i\pi\sqrt{1-4/t}}{\lambda}\ln\left(\frac{\lambda s_+
	-\lambda_1}{\lambda s_- -\lambda_1}
	\cdot\frac{\lambda s_- -\lambda_2}{\lambda s_+ -\lambda_2}
	\right)\ ,\\[12pt]
\rho_{a\pm}&=&\frac{(t_+ +1)t-6 \pm\sqrt{3(tt_+ -t+1)(t-4)}}{2(t-3)}\ ,\\
\rho_{b\pm}&=&\frac{-(t_+ -2)t-6\pm\sqrt{3(1-tt_+)(t-4)}}{2(t-3)}\ ,\\
\eta &=& \frac{(t-4)s+\lambda}{2(t-3)}\ ,\qquad
	\lambda=\sqrt{s} \sqrt{t-4} \sqrt{(t-4)s-4t+12}\ ,\\
\lambda_1&=&\frac{s(t-4)+\lambda}{2}-2t+tt_+ +6\ ,\qquad
	\lambda_2=\lambda_1+t-2tt_+ \ ,\\
\eta_a&=&-\frac{\left[(t-4)s-tt_+ -t+6\right]
	\left[(t-4)s+\lambda\right]-2(t-3)
	(t-4)s}{2(t-3)\lambda}\ ,\\
y_{1a}&=&\eta_a+\eta,\qquad y_{2a}=\frac{\eta_a}{1-\eta}\ ,\qquad
	y_{3a}=-\frac{\eta_a}{\eta}\ ,\\
\eta_b&=&-\frac{\left[(t-4)s+tt_+
-2t+6\right]\left[(t-4)s+\lambda\right]-2(t-3)
	(t-4)s}{2(t-3)\lambda}\ ,\\
y_{1b}&=&\eta_b+\eta,\qquad y_{2b}=\frac{\eta_b}{1-\eta}\ ,\qquad
	y_{3b}=-\frac{\eta_b}{ \eta}\ .
\end{eqnarray*}

	We have verified that the gauge and Higgs boson scattering
amplitudes obtained by using the results of this Appendix
satisfy the relationship (\ref{unit}) between the imaginary parts of
the loop amplitudes and the iteration of the tree amplitudes.
Ambiguities encountered by us in the evaluation of the imaginary parts
of Triangles~3 and~5 and Boxes~2, 3~and~5 were also resolved with the
help of (\ref{unit}).

\cleardoublepage

\begin{figure}
\caption{  Diagrams for the two-point functions. Tadpole-tree diagrams
are not shown as explained in the text.  In Figs.~\hbox{1-5}, solid
lines represent gauge bosons, dashed lines represent the Higgs boson,
and a cross represents a counterterm.  Also, bubble diagrams labelled
with $w,z$ represent two diagrams with $w$ and $z$ loops, while
triangle and box diagrams labelled with $w,z$ represent three diagrams
with two $w$~loops and one $z$~loop.\label{two.point} }
\end{figure}
\begin{figure}
\caption{ Diagrams for three-point functions. Lines on the left correspond
to external particles. }
\label{three.point}
\end{figure}
\begin{figure}
\caption{ Diagrams for $w^+w^-\rightarrow zz$ scattering. }
\label{wwzz}
\end{figure}
\begin{figure}
\caption{ Diagrams for $w^+w^-\rightarrow HH$ scattering.}
\label{wwhh}
\end{figure}
\begin{figure}
\caption{ Diagrams for $HH\rightarrow HH$ scattering.}
\label{hhhh} \end{figure}
\begin{figure}
\caption{ Absolute values of the S-wave
\hbox{$W^+_LW^-_L \rightarrow W^+_LW^-_L$},
\hbox{$W^+_LW^-_L \leftrightarrow Z_LZ_L$}, \hbox{$Z_LZ_L\rightarrow Z_LZ_L$},
\hbox{$W^+_LW^-_L \leftrightarrow HH$} and \hbox{$HH\rightarrow HH$}
amplitudes as
functions of the center-of-mass energy $\protect\sqrt{s}$ for
$m_H=$500~GeV. The \hbox{$Z_LZ_L\leftrightarrow HH$}  amplitudes,
not shown, are
$1/\protect\sqrt{2}$ times the \hbox{$W^+_LW^-_L \leftrightarrow HH$}
amplitudes.
The solid lines represent the Feynman amplitudes, the dashed lines the
K-matrix amplitudes, and the dotted lines the Pad\'e amplitudes. }
\label{amps1}
\end{figure}
\begin{figure}
\caption{ Same as Fig.~\protect\ref{amps1} for $m_H=1000$~GeV. }
\label{amps2}
\end{figure}
\begin{figure}
\caption{ \protect\sloppy The invariant-mass distributions for the
collider processes
$pp\rightarrow W^+_LW^-_L {\ \rm or\ } Z_LZ_L $ $\rightarrow W^+_LW^-_LX$ and
$pp\rightarrow W^+_LW^-_L {\ \rm or\ } Z_LZ_L \rightarrow Z_LZ_LX$
at the SSC energy of
$\protect\sqrt{s}=$40~TeV
for $m_H=$500~GeV and 1000~GeV. The dashed lines correspond to the K-matrix
amplitudes, and the dotted lines to the Pad\'e amplitudes.
A rapidity cut of $\eta_c=1.5$ is imposed on the
final $W$'s and $Z$'s. }
\label{cross}
\end{figure}
\begin{figure}
\caption{ Comparison of the K-matrix
\hbox{$W^+_LW^-_L \rightarrow W^+_LW^-_L$} amplitudes
without and with the one-loop correction for $m_H=$500~GeV and 1000~GeV.
The dot-dashed lines represent the tree amplitudes, while the dashed lines
represent the amplitudes with the one-loop correction. }
\label{compare}
\end{figure}

\begin{thebibliography}{99}
\bibitem{wznote} The real and imaginary parts of the Feynman
	amplitude (\ref{wz})
	for $w^+w^-\leftrightarrow zz$ are, of course, the same as
	those obtained
	in Refs.~\cite{DW} and~\cite{dicus}, respectively.
\bibitem{ampnote} The three bubble integrals, the first two triangle
	integrals and
	the first box integral have been evaluated by Dawson and Willenbrock
	in Ref.~\cite{DW}, while their
	imaginary parts are given in Ref.~\cite{dicus}.
\bibitem{spence} Following the usual practice, we have denoted the
	real part of the
	dilogarithmic function ${\rm Li_2} (z)$ as ${\rm Sp}(z)$.  However,
	for Triangles~5
	and~6, we have found it necessary to use the ${\rm Li_2} (z)$ notation.
\end{thebibliography}
\end{document}